# Quantization of the Atom plus Attempting to Answer Heilbron & Kuhn

Yeuncheol Jeong[1], MingYin[2] and Timir Datta[3]

1-Sejong University, Seoul, South Korea; 2-Benedict College, Columbia, South Carolina, USA; 3- University of South Carolina, Columbia, South Carolina, USA

**Key Words:** Atom, Quantum, Earnshaw's theorem, Induction Coil, X-rays, Laue Spots, Bragg, Moseley, Rutherford, J.J. Thomson, Nicholson, Quantization of Angular Momentum, Quantum Stability, Nobel Prize.

**Synopsis:**

The idea of atoms is old but X-ray provided the first probe into the physical atom. Photographs of X-ray scattering from crystals -'Laue spots'- were the first visual proof for the physical existence of atoms arranged in a perfect geometric pattern; thereby conclusively established the stability and physical reality of atoms. The father and on team of the William H. and William L. Braggs developed Laue's technique to study atoms and their aggregates. Inspired by Laue the hapless Moseley applied (Bragg's) X-ray spectroscopy to determine the nuclear charge number of Rutherford's atom and essentially settled a conundrum of the source of atomic mass that had baffled J.J. Thomson and stymied the plum pudding atom. In the past historians have quipped about the timing of Bohr's interest in Rutherford's atom. We argue that Bohr also at Manchester and contemporary of Moseley likely was inspired by Laue's discovery to get busy with the mechanics of the nuclear atom. Bohr synthesized extant ideas to formulate his own quantized theory of atoms and published a set of three papers the now famous trilogy one hundred years ago in 1913. Roentgen's discovery was awarded the first Nobel prize ever in 1901, Laue was honored in 1914, the Braggs in 1915, making Lawrence Bragg then at 25 the youngest ever. Eleven of the cited authors (Bohr himself included) in the trilogy; but not the most cited (John Nicolson), were later recognized by ten Noble prize awards, seven Laureates in physics and four in chemistry. The discovery of X-rays, the ensuing theoretical quantization of the atom and Nobel honors played a critical synergetic role in transforming the how and what of modern physics.

**Introduction:**

It was in a lecture at the Royal Institution in 1803 that Dalton first to put the atom concept into a logical and quantitative frame work. However many of the practicing chemists of the nineteenth century did not consider atoms to have physical reality even though they may apply Dalton's formulae in quantitative chemistry. The concept of atoms as the fundamental building block of bulk matter and developing a working theory of atoms took time. Affirmations of the quantum atom and the ever deeper quests into the structure of the subatomic realm have been central occupations of physics in the last hundred years. Although atomicity has a long history, it was only after the discovery by Roentgen in 1895 of X-rays that the atom could be meaningfully probed. During the first few decades of the 1900s, quantum ideas were developed to adduce the theory of the fundamental constitution of matter. In this article we explore some of the characters, background technologies and sequence of activities, which are all related to the first comprehensive quantization of atoms. The atom concept is so pervasive in our society's consciousness that most people would associate by sight figures of miniature solar systems with atoms.

In November 1911, the first Solvey council was held in Brussels. This conference represents an epochal moment in modern physics, it spot-lighted the conundrum posed by the new theory of radiation and the concept of the quantum. Nineteen extant experts from six European nations examined the most pressing problems of that era. In particular, Sommerfeld's theory of X-ray waves, as well as the pioneering contributions of Haas and Nicholson, was discussed. In the summer of the following year (1912), physics witnessed a chain of momentous events. Two discoveries with X-rays one by Laue at Munich and the other by Braggs in Cambridge and Leeds were direct cause and effect that inter alia landed two Nobel prizes.

Another story is the quantization of Rutherford's atom which culminated with Bohr's trilogy of articles in Philosophical Magazine, exactly one hundred years ago in 1913. The making and impact of the now famous trilogy entitled "On the constitution of atoms and molecules" will be discussed in this article. The trilogy directly contributed to a number of Nobel prizes including Planck's and indirectly to many others'. It might not be coincidental that the preeminence of the quantum in modern physics and the emergence of the Nobel Prize as the world's uber-symbol reflect a mutual symbiosis. Finally in answer to Heilbron and Kuhn's 1969 question - what inspired Bohr? We present a set of spatio-temporal evidence that relates Laue's discovery as the plausible trigger for Bohr to abruptly withdraw from his extant projects and get busy with his atomic theory.

**Road to the first building block:**

The concept of atomicity is very important in understanding matter and not surprisingly this idea has a long history. There are two distinct pathways, namely, the top down and the bottom up approaches. The bottom up scenario starts with the smallest possible constituent and builds upwards with increasing complexity toward our sensory macro scale. Albeit, accepting the de facto existence of such a first building block is in itself a logical challenge. Notice, that due to the finite dimension at the start this upward construction is a process of finite steps. Conversely the top down view starts from matter in the macro scale and proceeds to query progressively into ever smaller components. The non- granular top down process with infinite divisibility at any scale size leads to the concept of absolute homogeneity.

Democritus and his teacher Leucippus in the 5-4$^{th}$ century BCE in Greece are credited[2] with the idea of atoms as the smallest and indivisible building block of matter. However, the mentor of Socrates, Anaxagoras famously did not cater to the granularity idea and instead favored absolute homogeneity. He reasoned that practicality aside, at least as a matter of principle, it is natural that matter should be divisible ad-infinitum down to any scale and argued against atoms. He pointed out that the granularity idea is unnatural because it sets an artificial finite (lowest) limit to the divisibility and results in some object to which the concept of (further) partitioning does not apply.

But the modern ideas about atoms began with the grammarian and meteorologist John Dalton's 1808 opus magnum, *A new system of chemical philosophy*[3]. It took a bottom up chemist's perspective. Dalton explained the extant chemical knowledge in terms of aggregations or dissociations of indivisible atoms. Another milestone was the 1811 essay by Amedeo Avogadro[4]. He emphasized that the essence of all matter is inertia or mass. Avogadro was the first to realize that gases under identical conditions of volume, pressure and temperature can have different (mass) weights but have the same number of particles, honorably, the Avogadro number. Samuel Earnshaw's [5] 1842 article on molecular forces brought forward the theoretical considerations associated with mechanical stability and equilibrium. In simple terms, Earnshaw's theorem goes as follows; suppose a classical (no quantum mechanics yet!) system in which magnetic north and south poles or positive and negative electric charges are present. Such a system can be in static, stable equilibrium only when the poles or the charges are attached or stacked on top of one another. This theorem holds even when constant external forces such as gravity are included but breaks down when diamagnetic (quantum) components are present. Anyone who has played with bar magnets and iron nails will easily appreciate Earnshaw's theorem by observing that a magnet may easily lift up a nail (or another magnet) off the table so that the nail ends up sticking to (stacking on) the magnet, or make the nail stand up with tail still touching the table but it is impossible to freely hang the nail in midair for any length of time. Another mechanical analogy is the balancing a finely sharpened pencil on its sharp end. Even if one manages to put the pencil stand upright it will soon tip and fall over to a side. Not surprisingly this theorem is extremely important in determining stable (classical) configuration of any collection of disparate charges.

Another work is Dimitri Ivanovich Mendeleyev's invention of the periodic table of the elements in 1869. Dimitri was one of the first to notice the recurrence in physical and chemical properties of elements from the lighter to the heavier elements. This systematics permitted him to group all the known elements in a cyclic fashion into the "periodic table". In addition to nicely organizing the elements, this arrangement provided qualitative and quantitative information as well as predictive powers. The empty spaces (boxes) left in his table indicated elements yet to be discovered with anticipated chemical and physical properties that would fit the expected behavior of the box in question. Furthermore, the relationship between the element's atomic weight and the ordinal number of the element's place in the table provided further insight into the atoms. After the introduction of the nucleus in 1911, researchers could associate each element's position in the table with the number of positive charges (protons) in its nucleus.

Another notable invention was optical spectroscopy by Gustav Kirchoff and Robert Bunsen in the 1860s. Spectroscopy enabled scientists to uniquely identify pure chemicals from the wave length of the light emitted, for instance, when the substance is heated up to a high temperature in a Bunsen burner. J. C. Maxwell's classic article[1] in Nature, "Molecules" can be mentioned as an extremely readable accounting of the state of affairs until 1873. Johann Jakob Balmer's (1885) and Johannes Robert Rydberg's (1888) discoveries of regularities in the spectral lines of hydrogen were other indicators to systematic nature of atomic characteristics (wave length) of the light emitted. Both Balmer and Rydberg also derived compact empirical formulae that indexed the spectral lines by integer numbers.

In the late nineteenth century, the most influential theoretical developments on electrodynamics were by Maxwell[6] and his followers. Remarkably, for a long time, it was not clear as to what is the smallest part of matter; there remained confusion between molecule and atom. Maxwell cleared some of the ambiguity by stating that "An atom is a body which cannot be cut in two… A molecule is the smallest possible portion of a particular substance"[6]. It is worthwhile to remember that even in the twentieth century skepticism persisted as to the physical reality of atoms. The atom-molecule rivalry, that is, whether matter is a conglomeration of molecules or atoms continued well into the twentieth century.

Another question is - can the 'indivisible' atom itself be composite? Can the atom be made of opposite north and south magnetic polarities or positive and negative electricity? How to put them together seemed impossible in face of the works of Earnshaw, Maxwell and others. They identified insurmountable stability problems in electric or magnetically bound systems. For example, Earnshaw's theorem influenced the plum pudding model by J. J. Thompson[7] where the electrons (negatively charged plums) are immersed in a continuous distribution (pudding) of positive charges.

As with the radio wave's a decade earlier, the discovery of X-rays created a rage. Many scientists were attracted by its possibilities and eventually X-rays morphed into "a research tool" as Ian Hacking would describe[8]. It is important to note that this high energy radiation was also the first contact with the subatomic. In his book[9] *From X-rays to Quarks*, Emilio Segre Segre (the 1959 Nobel Physics laureate) was absolutely on target in pointing out the importance of X-rays. Really, a whole bunch of modern science emerged directly from Roentgen's discovery.

However, in 1900, at the British Association for the Advancement of Science, Sir William Thomson, aka Lord Kelvin, the supreme "electrician" in J C Maxwell's fabled treaties, famously proclaimed, "[t]here is nothing new to be discovered in physics now. All that remains is more and more precise measurement" [citation??]. Incredibly, in his assessment, William Thomson has conveniently ignored extant knowledge that since proven to be momentous; notably X-ray, radioactivity and electrons. The discoveries of electron (1897) and radioactivity (1896) causally followed that of X-rays.   In his attempts to generate X-rays from phosphorescent potassium uranyl sulfate, Henri Becquerel discovered radioactivity. Likewise while investigating ionization of gasses by X-ray, J. J.Thomson was able to isolate in 1897 for the first time ever the negatively charged elementary particle (electron). Incidentally, Thomson was using a modified version of Perrin's 1885 experiment on negative charges. It is unthinkable how the evolution of the physical sciences would be like without these three novelties, especially, without X-rays.

**Enabling technologies for X-ray:**

In his book, Image & Logic, Peter Galison[10] treats physics as a collection of activities in three "quasi-autonomous subcultures" of instrumentation, experimentation and theory. The discovery of X-ray in the late nineteenth century was an outcome of the instrumental developments in electrical, vacuum and photography technologies. Experimentations with this radiation in the first decade of the $20^{th}$ century provided the first glimpses into the submicroscopic and the additional momentum to solving the then hot issue, theory of the quantum atom. X-ray can thus provide a case study of these three subcultures.

Even instrumentation alone is a classic triptych of three ingredients; each was required, and equally essential in the discovery of X-rays. These ingredients are high voltage electricity, good vacuum and photo-plates.

First, to produce X-rays, a source of high potency (many thousand volts) electricity is a must. Roentgen used a Ruhmkorff induction coil as his source. The peripatetic inventor Heinrich Daniel Ruhmkorff was undoubtedly the most notable developer, promoter and marketer of induction coils. He took the best extant design practices and *ceteris paribus* built his own version. In particular, he incorporated the American design style of low-aspect-ratio long coil as opposed to the European large diameter short coil. This facilitated better flux linkage (high mutual inductance) between the primary and the secondary turns. He also retained the German innovation of the automatic electro-mechanical oscillator with an adjustable 'make & break' contact. Such automatic oscillating contacts permit high rate of flux change and hence inducing high voltage in the secondary over a range of frequencies. In addition, Ruhmkorff's own contribution was 'the shellac insulation' on the coil wire, so that the required high voltage can be maintained without electrical leakage. In 1858 he was recognized by Napoleon III with an award of 50,000-francs for the invention of most important electrical apparatus of that era.

Second, an electrically insulated and evacuated chamber with inside air pressure about a millionth of atmosphere is also needed for X-ray generation. This was an innovation from the 1870s principally by (Sir William) Crookes, the British chemist spectroscopist and inventor of the Crookes radiometer. He and his instrument makers introduced all glass (no rubber) bake-able enclosures with vacuum-tight-ground-glass seals. Preheating or baking allowed degassing (i.e., removal of gasses and moisture) from the interior surfaces of the apparatus. To reach the desired low pressure, the Crookes team employed 'mercury drip' vacuum pumps. These pumps were developed by Geilsler's team in Germany.

The final third ingredient is some means to record, measure and preserve evidence of X-rays. Dry photo plates served this purpose very conveniently. In 1871 Richard Leach Maddox, an English physician, is credited with the idea of using a dried layer of gelatin and silver bromide emulsion spread on a piece of flat plate glass. This is 'the dry plate' for photography. Prior to this development, a photographer had to be reasonably competent with 'wetlab' chemical processes. Shortly before exposure, the wet plate had to be sensitized by dipping it in a chemical solution. Then this wet plate has to be loaded into a camera for exposure. Quickly, this exposed plate will have to be put in another chemical solution to "develop" the image. The image is further "fixed" to desensitize in still another wet rinsing bath. The fixing process protects the plate from light. However, with the passage of time and light exposure, the image would tend to rapidly bleach and fade. Needless to say, all of the chemical steps were to be executed under no light conditions. Dry plates conveniently separated the chemistry of plate fabrication and image processing from a picture taker's hands. Dryness also permitted relatively long shelf life time before and after expossure. By 1878 such plates were commercially manufactured. Within two years of the advent of commercially available dry-photo plates, George Eastman started his own Eastman Dry Plate Company in 1880 and the (Eastman) Kodak was in business.

Naturally, many historians have already noticed the importance of both high voltage and high vacuum, but interestingly have remained relatively silent about the role of photography in the development of science, especially about that of the dry plate. Roentgen's discovery of X-rays took place in less than twenty years since the introduction of these plates. Such plates were critical in Becquerel's radioactivity discovery as well as in the famous specific charge (q/m) measurement in Thomson's parabola experiment. In this experiment, a beam of ions was subjected to 'crossed' electric and magnetic forces at right angle and hence get deflected into parabolic curves leaving characteristic traces on the photo-plate placed at the end of the gas discharge tube. From the photographs, the charge to mass ratio of the ions can be determined[a].

**The first Solvey council in 1911- radiation theory & quantum**

During the first decade of the 20$^{th}$ century, whether X-rays were waves or particles has no clear answers. Newer discoveries in other areas such as in radioactivity provided no better understanding, either. This created a big conundrum. To make sense of the state of confusion, a collective effort was called for - the first Solvey counclie[11], held in Brussels in November of 1911, Belgium. Borrowing a phrase from Max v Laue[12], this invitation only event was attended by nineteen "acknowledged masters" from six European nations including Curie, Einstein, Jeans, Onnes, Perrin, Planck, Rutherford, Sommerfeld, Wien and others. Table-I provides a list of the invitees with asterisk marks indicating the past and future Nobelists amongst them. They deliberated on the general topic of the ultimate constituents of matter, specifically, of radiation theory and the quanta.

The general importance of the Planck constant was discussed in the context of an action "h". It means Sommerfeld's work on X-ray Bremsstrahlung and h-hypothesis, contributions by Erich Arthur Hass from Vienna and John Nicholson from Cambridge were all examined. Arthur Erich Hass was the first to realize that Planck's constant was essential in fixing the atomic dimension and as early as in 1910 computed the size of an atom incorporating "h". At the conference, his estimated value of ~ $2.8 \times 10^{-10}$ m as the size of the (then current) Thomson's model was reported[13]. In 1911, Sommerfeld had an assistant, Walter Friedrich, a former Roetngen PhD, to experimentally test his h-hypothesis. Unfortunately, in November of 1911, the experiment was still dragging along. Hence, in answer to a query by Lorentz at the meeting, Sommerfeld paraphrased Goethe's famous line of "planting roses without knowing if they will bloom", implying that blossoms are not ready yet! This remark emphasizes importance of experimental results and should be a timeless reminder to all of us.

This first Solvey meeting remains as a landmark in history of physics, where arguments for and against the usual forms of the dynamical laws were debated. In addition, the frisson of the novel and the associated risks were noted. Rrecent successes of quantum ideas were reported from radiation theory. The pesky failures of classical ideas such as Dulong-Petit law of specific heat were expounded. Noticeably, James Jeans asked a question as to whether this new [quantum] method was capable of "providing an image of reality". His question was raised well before the emergence of the full quantum description including uncertainty principle, wave-particle duality, wave function, spin and others; Sir James was at least one and a half decade ahead of his peers. In many of the future Solvey meetings, the topics on quantum reality would be prompted by Einstein through his various thought experiments. Even to this day, the foundations of quantum mechanics have remained as an exciting area of investigation, thanks to the efforts by some others such as our erstwhile colleague Yakir Aharonov.

**X-rays in UK & Germany:**

During first decades of the twentieth century in the UK, the two dominant and influential experimental physicists were the Nobel laureates J.J. Thomson (known as J.J to those familiar with him) at Cambridge and Earnest Rutherford at Manchester. Also important to our story were the radiation researchers, William Henry Bragg, Charles G Barkla and Henry Gwyn Jeffreys Mosley. Henry Bragg was the first to perform X-ray experiments in Australia and after 22 years at Adelaide University had returned to UK in 1909 to join the University of Leeds. Rutherford had already visited the Bragg family at Adelaide in 1895 and have kept in touch with them since. At Cambridge, based on observation of polarization, Barkla had discovered the transverse nature of X-radiation in 1906. Moseley at Manchester, later in the autumn of 1912, started to investigate X-ray diffraction and finally developed his eponymous law of X-ray spectra. This promising young man was expected to have a great career but unfortunately Moseley ended up being killed in battle (August, 1915) at Gallipoli during WW-I.

However, there was hardly any British consensus on X-rays. For example, a particulate model of X-rays comprising of neutral pairs of opposite charges was hypothezied[14] by Henry

Bragg. He used cloud chamber images to bolster the idea. The fact that both electrons and X-rays arise from the same source, namely, thru high voltage discharges in rarefied gasses, supports the corpuscular perspective. The difference between them is that electrons cannot get out of the glass enclosure, whereas X-rays are far more energetic and penetrating. Furthermore, once out, X-rays continue to pass thru other barriers as well. On the other hand, Thomson regarded[15] X-rays as localized "bundles of energy". The results of Barkla's scattering experiment with X-rays[16] seemed to follow an angular dependency anticipated in Thomson's model. However, J. P. V. Marsden's gamma ray results[17], indicating more bias towards scattering in the forward direction, cast a doubt on the energy bundles idea. The lack of conclusive evidence for or against particles eventually led Lawrence Bragg to suggest a sort of dual "wave-particle" nature to X-rays. Unfortunately in 1910 Bragg's pioneering duality concept was too novel and did not get much traction. Prior to Louis de Broglie's introduction of wave-particle duality in 1927, waves were supposed to be waves, and particles just particles.

At the heels of the discovery of Hertzian waves, amongst many researchers in Germany, the sentiment about X-rays favored a wave model. For instance, as early as in 1896, Emil Wiechert and George Gabriel Stokes have argued, based on the way of X-rays generation, that this radiation must be short waves of electromagnetic pulses. Arnold Sommerfeld was one of the champions for the wave theory of X-ray. In 1900, at the request of the Bavarian government, Wilhelm Conrad Roentgen moved to Munich to become the professor of experimental physics a position previously held by E. Lommel. Roentgen in turn recruited Arnold Sommerfeld to Munich in 1906. Sommerfeld, a successor of Luwig Boltzmann (who had left Munich for Vienna in 1893), became the director of the newly established theory institute at Munich. Roentgen and Sommerfled exercised a combined effort in solving the "nature of X-rays". As a consequence, Sommerfeld's theory team got its own laboratory facility in the basement of the institute, where people were mostly experimentalists with experience under Roentgen. From his measurement of photoelectric effect in 1907, Wilhelm Wien estimated the wavelength of X-rays to be $7 \times 10^{-11}$ m. In Sommerfeld's group, the wave length of X-rays was measured from fringe patterns produced by steel edges. Their estimation was $4 \times 10^{-11}$ m, the best (extant) value until 1912.

**Max v Laue:**

Starting in 1909, Max (not yet a von!) Laue became a lecturer in optics (*Privatdozent*) for Sommerfeld. Laue had completed his PhD with Max Planck on thermodynamics of coherent radiation and was already well known for his encyclopedia article on optics. In the summer of 1910 Paul Ewald entered Sommerfeld's group and selected the problem: 'To find the optical properties of an anisotropic arrangement of isotropic resonators'. During the time between the Christmas break of 1911 and January 1912, Ewald had finished calculations and, writing a thesis, he told Laue about his results. Laue appeared curious about possible behavior of very short wavelength radiations. Ewald evidently had aroused Laue's interest enough to test the wave nature of X-rays by crystal diffraction[18]. As Laue would explain[19] in his Nobel lecture (1914), he was motivated to learn " … which diffraction effects with X-rays might be found, and [whether] the question of their true nature [can be] answered …".

While in the spring of 1912, Laue might not been aware that Roentgen himself had already studied the scattering of X-rays by crystals shortly after the discovery, but in vain; "I [Roentgen] continued the experiments to which I referred already in my first communication about the transparency of plates of equal thickness that have been cut from a crystal along different directions, Again, no influence of direction on the transparency could be recognized" – without the slightest indication of a diffraction effect[20].

However, unlike the simple discharge tube that was at Roentgen's disposal, by 1912, dedicated x-ray tube design and target technology have made it possible to generate higher intensity and arguably better collimated beams by the folks at Munich. Now, the diffraction experiment was possible by Laue. As some authors question, how could Laue, a theoretician, be so bold enough to suggest such an experiment? There is no direct archival record – in the form of letters, diary or manuscript – from which Laue's motivation would become clear[20].

Nonetheless, the story leading to Laue's idea goes as follows. On a skiing expedition during the Easter break of 1912, Laue seemed to propose an idea on X-ray diffraction experiments with crystals. Sommerfeld, Wien and others promptly cast doubts on the proposed diffraction pattern which, according to them, can be scrambled up by atomic zero point motion, random vibrations of atoms that persists even at absolute zero temperature. Peter Debye and Ivar Waller later developed[21] the theory on zero point contributions, showing that spots would nonetheless survive but with reduced intensities. This now famous Debye-Waller treatment has formed a basis of the standard description for many scattering processes such as the Noble winning Mossbauer processes.

Back in Munich, not taking no for an answer, Laue resorted to what was described in his Nobel lecture as "a certain amount of diplomacy". According to Ewald, Laue brought the issue with the regulars on the famous "physics table in the Café Lutz", where the consensus was finally formed that Laue's experiment was worth a try although his theory was a bit shaky. Then, with no elaborate set-up needed, the Laue project did not seem to interfere with the schedules of the Sommerfeld–Friedrich experiment in progress (as noted earlier). Before the end of April, after some initial setbacks[22], by directly passing the X-rays thru a large copper sulfate crystal, Laue's two associates, Friedrich and Knipping succeeded in producing the interference patterns and took pictures of the "Laue spots". In the photographs by Freidrich et al., the X-ray diffraction spots appeared to qualitatively follow Laue's original expectations, but the details were not quite right. Furthermore not all the predicted spots were seen[23,] either.

**Importance of Laue's pictures:**

The spots were shown to have been produced by the X-rays of the same wavelength as the incident radiation. Hence, the pattern was not due to secondary (characteristic) fluorescent X-rays re-emitted by the radiated atoms, as Laue had initially proposed. Also, nor can they be formed by 3-dimensional diffraction due to the crystal lattice. Instead, as was soon to be proven by Lawrence Bragg, the spots arose from the interference of X-rays which are reflected off layer-by-layer from two-dimensional atomic planes inside the lattice. Other researchers elsewhere also took note of these discrepancies. For example, Moseley, who in the autumn of 1912 undertook his own X-ray diffraction experiments at (Rutherford's) Manchester, wrote to his mother that the Munich group did not fully understand what the results meant and their explanation on the spot formation was not right either. Thus, Laue's initial understanding was questioned. But, to his credit, Laue quickly recognized the shortcomings in his initial explanations. For instance, when in 1914 the Nobel committee cited his achievement *"for his discovery of the diffraction of X-rays by crystals"*, Laue corrected his own Nobel lecture title as "*Concerning the Detection of X-ray*

*Interferences*". The change from "diffraction" to "Interference" was no typo! This reflects Laue's embracement of the questions raised by Bragg, Moseley and others.

However, one should not underestimate Laue's strength of creativity! Deep down in Laue's gut, he had felt profoundly unsettling and continuing questions. Laue sensed an expanded vista of new insights. Generations of scientists will follow many of these unexpected and far reaching questions. Since his discovery, over a dozen Noble prizes have been awarded which can be directly traceable to Laue's work. Even today, X-ray spectroscopy and crystallography continue to be valuable in many fields including chemistry, geology, medicine, molecular biology, materials science and physics. In the context of Laue's tremendous influence, Peter Debye[24] once remarked that "one should generally not trade merit against luck with such things".

It is interesting to note that the roles of the objects are completely opposite in Roentgen's vis –a- vis Laue's pictures. In the first case, the picture of Frau Ronetgen's hand served to confirm the discovery of penetrating radiation. Roentgen's picture relied on the firm established medical knowledge of orthopedics to validate the discovery of this hitherto unknown radiation. Everyone concerned was absolutely sure about the skeletal structure of the human hand. Thus, the object was nothing new in the image of these particular bones. By contrast, the initial motivation of Laue's experiment was to confirm the wave nature of X-rays which got relegated to be incidental. In Laue's photograph, instead, the objects in the picture, the atoms, took over the central role. Notice that, in the Roentgen picture, the image of the wedding ring played only a subjectively important role. The ring subliminally vouchsafed the veracity of the discovery. On the other hand, in the Laue picture, the perfect and rigid geometric beauty of the spot pattern manifested an axiomatic certainty for the atom's stable existence, a sort of a proof by geometric arguments typically reserved for the theorems such as in Euclid's Elements.

The tremendous significance of Laue spots was clear right away. Even before the publication of the reports, Walter Friedrich, Paul Knipping and Max Laue realized it and agreed to sign a one-page document stating "The undersigned are engaged since 21 April 1912 with experiments … of x-rays passing through crystals". It was officially deposited by Sommerfeld[20] as head of the group on the 14$^{th}$ of May in 1912. The first publication[25] was by Max Laue, W. Friedrich and

P. Knipping. Although all three names appeared in it, two years later in 1914, the Nobel Prize in physics was awarded only to Max von Laue alone. Laue felt that his two co-workers should have shared it with him and simply chose to distribute the money equally between the three.

**Senior and Junior Bragg, the only father and son Nobel Laureates**:

The news of Laue's discovery spread rapidly arousing great curiosity and sensation everywhere. While vacationing on the Yorkshire coast with family, Henry Bragg received a long letter[26] about Laue's results from Lars Vegard, a friend of Henry's and a former visitor to his Leeds laboratory. Vegard was a Norwegian aurora physicist and at that time a co-worker of W. Wien at Wurzburg (in Northern Bavaria). He had attended a lecture given by Laue at Wurzburg. The letter sent on the 26$^{th}$ of June in 1912 with a copy of one of the experimental photographs, detailing Laue's findings. The senior Bragg discussed this with his prodigal son, Lawrence Bragg. At that time, Lawrence was only a twenty-two year old physics student at Trinity College, Cambridge University. Henry Bragg succeeded to get Lawrence interested. Coincidentally, both of his two sons Bob (another WW-I casualty; killed in the battle of Gallipoli, 1915) and Lawrence were students at St Peter's College, a school in Adelaide. St Peter's is the second only to the Bronx High School for Science in terms of the number of Nobel Laureates among its old boys.

As will be evident shortly, father's telling of these remarkable details must have extraordinarily inspired this twenty-two year youth into a creative burst. To start with, the Braggs considered Henry's 'neutral-pairs of charged particles' to be consistent with the spots. The spots can result from channeling instead of diffraction as proposed by Laue. Channeling is not a wave behavior, but the passage of the particle thru the hallways in the lattice. This mechanism can give rise to spots in different directions as the neutral-pairs traverse the crystal. The Braggs were pioneers because since the l970's ion-channeling and Rutherford-back-scattering have become extremely powerful tools in studying the distribution of defects in pure crystal solids.

Briefly after returning to Cambridge at the start of the school year, but still before Christmas, Lawrence not only confirms Laue's report, but also improves on the technique. The junior Bragg noticed a number of intriguing details[27]. First, not all the diffraction maxima predicted by Laue's fundamental equations actually appeared in the photographs. Incidentally, the Munich group had tried to explain away the absent spots by invoking ad hoc spectral distribution of the incident X-ray. Second, the shape of the spots changed from circular to oval as the pattern moved away from the direction of incidence. Third, crystal rotated by a certain angle results in the whole spot pattern's bodily turning by twice the angle. Typically, angular displacements of 'non-specular' interference spots are determined by a set of trigonometric functions which produces a highly nonlinear response[28], not the simple proportionality between the angles.

Within a few months, Lawrence succeeded in proving that the actual process for the spot formation is not by unrestricted three-dimensional diffraction, as Laue thought. But it is in reality due to the interference of primary X-ray waves that are reflected layer-by-layer from two-dimensional atomic planes inside the crystal lattice. By December of 1912, Lawrence published a paper in Nature describing the correct physical interpretation along with the derivation for the now famous Bragg formula. In this 'specular reflection' model, there are two relevant lengths (the x-ray wave length, the inter-plane distance) and one angle (between the incoming and outgoing waves). Using these variables, Lawrence also derived the appropriate conditions for constructive interference and obtained a simple eponymous Bragg formula. This equation relates the order of the interference multiplied by the ratio of the two relevant lengths with the trigonometric sine function of half the angle. In his 1915 Nobel lecture (delivered after WW I in 1922), Lawrence states " … I tried to attack the [Von Laue] problem from a slightly different point of view, and to see what would happen if a series of irregular pulses fell on diffracting points arranged on a regular space lattice. This led naturally to the consideration of the diffraction effects as a reflexion of the pulses by the planes of the crystal structure".

Furthermore, based on his equation and the missing spots in Laue's photographs, Lawrence deduce that the atoms in zinc sulfide (ZnS) are not in a simple cubic (SC) structure, as previously presumed by Laue, but in fact are ordered in a face centered cubic (FCC) arrangement. Not coincidentally, Sir William J Pope, professor of chemistry at Cambridge University (formerly at Manchester) got interested in William Lawrence's activities. Pope had developed the "valence-volume theory" of crystal structures together with William Barlow. Barlow, a noted amateur, was the first to introduce the concept of close packing as well as SC and FCC structures in the study of crystals in 1883. Also, Barlow had correctly guessed the atomic arrangement for the alkali halide crystals. In Lawrence's photographs, Pope saw a confirmation of his valence-volume idea and procured several specimens of sodium and potassium chloride crystals from Germany to provide further X-ray studies for Lawrence. People who have known the Braggs in person[30] have remarked about the individual contributions of father and son. Here, let us directly quote the sons own comments;

"it was the help which I got from Professor Pope, the Professor of Chemistry, and the inspiring influence of C. T. R. Wilson, which led to my analysis of sodium chloride and potassium chloride by the method of the Laue photograph. Pope and Barlow had developed a valence-volume theory of crystal structure, and when my first studies of Laue's diffraction patterns led me to postulate that zinc sulphide was based on a face-centred cubic lattice, Pope saw in it a justification of his theory and urged me to experiment with sodium chloride and potassium chloride crystals which he got for me from Steeg and Reuter in Germany. These experiments were made in the later part of 1912 and early in 1913. Simultaneously, my father seized on the conception of the reflection of X-rays by crystal planes to design his X-ray spectrometer, and discovered the X-ray spectra".

Lawrence did his own experimental work initially at the Cavendish Laboratory at Cambridge and, in his early papers, gave Trinity College for address. Shortly thereafter Lawrence left Cambridge to join his father at Leeds. From then on, many of their developments made at Leeds were of applied nature such as apparatus making, invention of x-ray spectroscopy and others. Arguably, at Leeds, the situation can be best described by a phrase from P.W. Anderson[29] "theory on tap not on top".

The father and son duo remain unique in many respects. They are the only father and son to share the same Nobel prize (1915 in physics) and Lawrence is the youngest (25 years) Nobelist yet! In the *grosso modo*, X-rays provided the first proof for the validity of the Planck-Einstein energy frequency (wavelength) relation in the high energy regime of radiation, extending the spectroscopic frequency laws to the K, L & M radiations of X-ray. They also provided the most important confirmation on the absolutely fixed geometric lattice arrangement of atoms, in other words, the absolute visual proof for the existence of stable atoms.

Amazingly, the concept of solid matter in a lattice structure of atoms (not in free molecules) did not sit well with the chemical establishment of that generation. For instance, in 1927, a former President of the (Royal) Chemical Society, H. E. Armstrong opined[31] in an article in Nature that "chess-board pattern of atoms" in NaCl (sodium chloride) as "repugnant to common sense... It is absurd to the $n$th degree... Chemistry is neither chess nor geometry ..." . Yes, it is odd to find out that crystals of sodium chloride, the common salt are not in a crystal arrangement of (NaCl) molecules, but rather in that of separate sodium and chlorine atoms.

**Rutherford 's atom:**

In 1911, based on a large number of experimental evidence collected in his laboratory, especially from the rare but important observations on alpha particles back scattering from thin foils of metal, Rutherford proposed a new atomic model which goes as follows[34]

"…Consider an atom which contains a charge $\pm Ne$ at its centre surrounded by a sphere of electrification containing a charge $\mp Ne$ supposed uniformly distributed throughout a sphere of radius… Taking R of the order $10^{-8}$ cm… The question of the stability of the atom proposed need not be considered at this stage, for this will obviously [*be true that stability of*] the atom is supposed to depend upon the minute structure of the atom, and on the motion of the constituent charged parts …".

Soon, the model was refined - the ambiguities with charges were resolved. In the new scenario negatively charged electrons orbit around a tiny and massive positive nucleus. But, even the refined model had detractors due to acute problems of stability associated with segregated and localized stationary charges. If an equilibrium is attempted by accelerating the less massive charge into an orbit around the heavy charge in a sort of a miniature electrically charged solar system, then the stability fails by virtue of energy loss from accelerated charges thru Maxwellian electromagnetic radiation. This classically enormous rate of radiative loss puts a severe limit to atomic longevity or the length of time. Such a system can only survive before the opposite charges collapse upon to each other.

**One John Nicholson, M.A.,DSc.**

Within about four months of Rutherford's publication of his first paper discussing the nuclear atom[34], John William Nicholson then at Trinity College, Cambridge, incorporated Rutherford's model. This is the same Nicholson[42] whose work was noted in the Solvey conference of 1911. Nicholson was amongst the early pioneers who regarded the condition of quantization to be essential in atomic mechanics. He was a maverick in his own right. Nicholson published some of his early results in an article[41] about "nebulium" in November 1911. Mindful of the perceived shortcomings of the nuclear model, he performed classical electrodynamic stability analysis under various perturbations on the rings. He computed the orbital frequencies of electrons and related them to the frequencies (wave lengths) of spectral lines observed in nebulae. The analysis is classical but sophisticated enough to reflect his awareness of recent advances in the theoretical physics such as the special theory of relativity.

In another paper[43] (submission date April 28, 1912) published in June, Nicholson extended the idea of quantization into a new realm. He reasoned that the principle of quantization on black-body radiation (an action associated with the field) should be more general and must include a mechanical action associated with particles in his 'atomic system'. Pursuant of this line of reasoning Nicholson showed that the ratio of kinetic energy to orbital frequency turned out to be exactly the orbital angular momentum which is also an action. Consequently, Nicholson argued that the permissible changes in "the angular momentum of an atom can only rise or fall by discrete amounts"[679-3$^{rd}$]; in other words the angular momentum of the atomic system is quantized. Remarkably, this insight has been missed by other experts in that era including Planck himself and Einstein. NB: Furthermore, such quantization automatically confers stability to the system, because by this quantum restriction the atom is prevented from gradually losing (/gaining) momentum (/energy) and decay (/rise) to some other state. Incidentally Nicholson's results were also the inspiration[44] for Wilson-Sommerfeld's quantization condition later on.

Demonstrating that "the ratio of energy to frequency" of a quantum system is an action, he directly deduced the quantization condition in angular momentum. Thus, he was the first to propose the condition of discrete jumps for the changes of angular momentum in a quantum system, although this change of angular momentum does not yet induce radiative emissions. He derived this proposal mathematically and logically without invoking any postulation or "special hypothesis". Incidentally, Nicholson chose not to assign a new name for the unit change of this action (angular momentum). If he did, perhaps we would call it Nicholson's constant not the reduced Planck constant or "h-bar" ($h/2\pi$). Such a difference in nomenclature is not uncommon. A case in point will be Avogadro's constant (number) and its cousin Loschmidt constant, these two are closely related, but not exactly the same. Hence, they bear different names. We observe that arguably the identification of "h" with angular momentum may be Nicholson's single greatest contribution to quantum physics.

**Niels Henrik David Bohr:**

Let us pick up the other summer happenings in 1912, this time at Manchester and our protagonist is the young Danish physicist, Niels Henrik David Bohr, who in the spring of 1912 has moved to the Victoria University in Manchester to conduct post-doctoral studies under Rutherford. The previous year in May of 1911, Bohr had completed his PhD in Copenhagen and had eagerly traveled to Cambridge to be mentored by J. J. Thomson. But after several unsuccessful encounters with Thomson, Bohr decided to end this unrewarding sojourn at the Cavendish. Although Rutherford was not yet as accomplished as Thomson, he was receptive to the ideas of Bohr.

At Manchester, Bohr befriended George de Hevesy, a Hungarian nobleman. Coincidentally, Hevesy's project at Rutherford's laboratory to chemically isolate the Radium D isotope from radioactive lead did not work out. This experience led Hevesy to develop the 'radioactive tracer technique'. This technique allows a researcher to keep tab on chemical reactions including biological process such as in plants and animals. He received the 1943 Nobel Prize in Chemistry *"for his work on the use of isotopes as tracers in the study of chemical processes"*. The prize was given one year later in 1944, presumably because there were no strong enough candidates in the original 1943 nomination docket. Bohr and Hevesy were to remain close friends for life. Perhaps, Hevesy was influential in drawing Bohr's interest to molecules and chemistry. Hevesy was socially well connected and introduced Bohr to Rutherford's laboratory in Manchester.

Bohr[32] was supposed to perform radioactivity experiments in the laboratory together with some theoretical work. The latter is related to some earlier work of the resident mathematician

C.G. Darwin regarding the absorption and scattering of "alpha and beta rays" and their transit time in matter. Bohr had some criticisms of Darwin's treatment and was hoping for a quick short note for publication in the Phil Mag. During May and June of 1912, he wrote candidly to his precautious younger brother Harald (mathematician, already with a PhD ahead of Niels!). For instance, in May 27, Bohr wrote about electron theory, noting a wrong order of magnitude in Thomson effect, lower specific heat of metals at low temperatures and high electric conductivity in metals etc. In another correspondence dated July 12, he renewed his thoughts on treating the electron theory. This letter included a lot of detailed mathematical equations associated with the absorption calculation. He promised to send Harald some more calculations to look over. The style of this letter was rather long, open and candid. Later in the summer, Bohr then signed off a manuscript on 'velocity of moving electrified particle' at Manchester in Augusts of 1912. The article appeared in print the following year (1913) in the January issue of Phil Mag[33]. Bohr's notable conclusion in this paper was that a neutral hydrogen atom has one negative and one positive charge. However, this paper does not relate to his famous theory of atoms in the trilogy paper.

**The Over Researched June/July 1912 Memorandum of Bohr:**

Curiously, by the third week of June, Bohr has stopped going to the lab. It seems that in this short time of June and July, Bohr has become preoccupied with Rutherford's atom. In the now famous[32] "June/July 1912 Manchester memorandum" made for discussions with his mentor Rutherford, he gives particular details concerning the (future) atomic model. Long gone are the topics of electron theory, with no mention of low temperature specific heat and not even a word about "velocity of Electrified particle passing through matter". No Sir! It was a dramatic new start with a big departure for Bohr, based on what he might have gathered at the Cavendish. The momrandum is all about atoms and molecules by themselves. Hence, the memorandum[32] is replete with corpuscles and full of sundry sketches on putative electrons rings circulating around. He recounted Planck and Einstein's quantization of radiation and then went on to the relationship between the energy and frequency of the radiation. He also took note of characteristic Roentgen rays and (Henry) Bragg's law of X-ray absorption.

In this memorandum, Bohr implied that the (final) stability is some type of 'statistical equilibrium'. By equating the thermal energy with the (classical) electrostatic potential energy of two elementary charges separated by hundred millionth of a centimeter, he then computed the temperature required to dissociate a "diatomic molecule". This classical estimate came out to be of the order of $10^5$ or around hundred thousand degrees (in Kelvin). Dissociation temperature of 100,000 degrees is incredibly too high rendering improbable stability, but it seems that he was not bothered by this large number. It will be interesting to know why?

Next, he attempted to compute the same dissociation temperature, from a different direction, by equating the thermal energy with the (quantum) energy of three petahertz (petacycles) photons. At this point, he also computed the value of the ratio between Planck's constant and Boltzman constant to be only one percent of the correct value. As a result, he writes "and this gives again $10^5$". However, if Bohr has used the correct values of the ratio from the two constants, then this would be around $10^3$ degree, that is hundred times smaller than and definitely not the same as his classical estimate.

After all, Bohr invented or postulated a special ad-hoc "hypothesis" for stability. This hypothesis is to become the corner stone of Bohr's quantum theory of the atom. In the memorandum, Bohr describes the hypothesis as "…will be a definite ratio between the kinetic energy …and the time of rotation". Shortly thereafter, he rewrites a mathematical equation, namely $E = Kf$. It seems, at this stage, K is just a constant of proportionality and he did not mention that this K is an "action". However, as we discussed earlier in the context of Nicholson's theory of atoms, this action is the angular momentum which Nicholson has already discretized, bestowing stability of the quantum state. Nicholson's result appeared in the June (1912) issue of the Monthly Notice of the Royal Astronomical Society; in this (June-July 1912) memorandum Bohr is either not aware of or has not yet fully grasped the implications of Nicholson's work.

**The Trilogy:**

Except for the last equation, most of the calculations and sketches in the memorandum did not appear in the actual publications, now famously called the Bohr-trilogy published in three parts in Phil Mag between July and November 1913. The last equation essentially restated the Virial theorem; i.e. for any stable bound orbiting system the potential energy for an inverse square law force is twice (with negative sign) the kinetic energy. Bohr applied this to obtain the first equation in the part -1 of the trilogy[35], "On the Constitution of Atoms and Molecules".

In this first paper (part-1) of the trilogy, Bohr references Einstein's contributions, emphasizing Planck's theory in the operations of atomic systems. From the physics point of view, Bohr system comprises of two distinct parts- the atomic and the radiation field and this system as a whole is isolated from the rest of the universe. Arguably, this partitioning is one of the most important conceptual steps. Another breakthrough was his determination of Rydberg's constant entirely in terms of the fundamental constants, namely, in terms of mass, charge and Planck's constant.

Bohr applies the quantization condition to both parts (atom and radiation) and stipulates that the principle of energy conservation must hold for the total system; i.e., the total energy of the system does not change with time. The total energy is equal to the energy of the atom plus the energy in the radiation field. Note that the principle of energy conservation dictates that as the system evolves in time the energies of the atom and the photon may change separately but the total energy remains the same. From the postulated stability hypothesis, Bohr showed that the stable atomic energy is inversely proportional to the square of an integer "n", the (principal) quantum number.

In Bohr's model, the frequency of the radiation is determined solely by the energy difference between the states of the atom. The ratio between the energy (difference) and the frequency is Planck's constant. From his expression for the energy, Bohr correctly reproduced the lines of the hydrogen spectra, thus providing a quantum mechanical derivation for Rydberg and Balmer's empirical relations. Bohr's formulation extended beyond visible spectra covering for both higher and lower energy radiations.

Notice also in this description, both emission and absorption of light (radiation) are possible - no qualitative difference. Einstein's photoelectric effect paper in 1905 dealt with absorption, whereas Planck's 1900 paper pertains to emission and so did Nicholson's papers. Bohr's was the first to include both processes, i.e., quantized radiation and quantized absorption. In Bohr's model, if the quantized radiation field loses energy, then the atom must gain that much energy. In other words, the atom has to absorb a photon to gain the requisite energy. Conversely, when a photon is emitted by the atom, then the atom has to lose its energy as the radiation field gains the same energy.

Bohr completed his three part opus magnum in rapid succession with part two on multi-electrons subtitled "Systems containing only a single nucleus" and with the last part on molecules, "Systems containing several nuclei". But, the results of these two later papers did not agree well with experimental data. In hind sight, even for systems with only one electron, the model did not give accurate results for the individual states but computed the energy differences with far better accuracy. Also the model had difficulties in explaining spectral intensities, effects of magnetic fields, the fine structure and the hyperfine structure of spectra lines.

The work of trilogy reflects the resources that Bohr had drawn upon; Rutherford clearly was the single biggest influence both scientifically and personally in providing with a smooth publication process. Bohr was very meticulous in referencing the contributions of his peers. Eric Arthur Hass's computations of the size of Thomson's atom and its connection with Planck's constant were duly credited[36]. In (X-ray) spectroscopy, the contribution of Barkla[37] was noted. Also, in the pesky question of atomic number and weight, the very recent work by van den Broek[40] was included and so was Bohr's own conclusion from the January 1913 article[32] about a neutral hydrogen atom. This atom was described to have only one negative and one positive charge.

In the entire trilogy the oldest article cited is one from 1896 (Part-1). In total, the number of distinct citations is thirty three. Bohr was extremely focused and he referenced especially the then recent literature; twenty two were only a year old (from 1912), and fifteen were from the same year (1913). As will be described shortly, this set of articles would become one of the most impressive and influential landmarks in physics. Reportedly, both Sommerfeld and Einstein were very enthusiastic about Bohr's new theory. But, not everyone was so positively inclined. For example, Otto Stern and Max v Laue who (although they later become strong followers) noted to have remarked that " if by chance it should prove correct then they [Stern & Laue] would quit physics" [Segre,p129,91980) ]. Nicholson also disagreed with Bohr and remained as a vocal critic [Kragh, Phys in Pers 13,p4-35(2011)].

**Citations in Bohr's trilogy:**

We reason most available matrices such as "citation index" or "impact factor" and others similar measures are only *a posteriori* indicators that assay after-trends amongst the readership. Such measures don't seem to reflect the true strength of the trilogy. To appraise the "uber" impact of Bohr's articles, we tracked the post-trilogy recognition of the authors cited in the articles[7]. Bohr showed an extraordinary skill in picking the most relevant and up-to date literature. The publication dates of the cited works range from 1896 to 1913.

As the reader will find in Table II, the trilogy referenced sixty seven works by thirty three different authors. The two papers by E.C. Pikering [in Astrophys J, IV, p369, (1896) and p92 (1897)] were the oldest references. Nicholson's relevant results were already in print by the time of the Manchester memorandum and more his details followed during the time of Bohr's writing. Bohr had met Nicholson in Cambridge. It is conceivable that Bohr perceived some insight from Nicholson's results. Also, Bohr might have checked his own work against Nicholson's. In many critical issues, Bohr pointed out where his own results agree with Nicholson's. In Part I of the trilogy alone, Nicholson is cited profusely, seven times to be precise!

Nicholson was the most mentioned (in parts I & II only) with a total of nine citations. Thomson was the second with seven and was the only author (other than Bohr's own self-citations) cited in all three parts. Rutherford was referenced five times, followed by Einstein's four and Planck's three citations. Table II gives a complete breakdown of all the authors cited in all three parts of the trilogy.

**Bohr's Quantum and Nobel:**

A quantitative measure of the tremendous impact of the trilogy is indicated by the large number (eleven) of post trilogy Nobel awards among the cited authors. In Table II, the cited names marked with two (**) indicate Nobel Laureates with awards prior to the trilogy and the twelve others who received a total of eleven Nobels after 1913 are shown with one asterisk. The Midas touch of the trilogy is quite phenomenal. For instance, Millikan's value of the electronic charge (in part-II) and Langmuir's result (in part-III) for heat of hydrogen [molecule] formation were both carefully noted by the award committee. Millikan received his Physics Nobel prize only a year after Bohr. Hevesy and Langmuir's Chemistry Nobels will follow, too. Curiously, Nicholson, the most cited author in the trilogy, was not recognized. Was he ever considered for the prize? The archives of the Nobel organization may hold interesting traces on how much, if any, influence the trilogy papers had in the deliberations for the prizes during that time.

Particularly, one archival record is especially noteworthy, these are the remarks regarding Bohr's contribution during Nobel presentation speech for Planck given by the Nobel Academy president A.G. Ekstrand; "still greater triumph was enjoyed by Planck's theory… [and] spectral analysis, where Bohr's basic work [was used]". Remarkably, here, Bohr appears to have become the fulcrum for Planck who was the founding father of quantization. Curiously, this is four years before Bohr himself gets his own Nobel recognition in 1922. A number of additional Nobel lectures of that era also directly referenced the trilogy; for instance, a year before Bohr's prize, Stark's mentioning of "Bohr's hypothesis". Then, again, in 1921, in his Nobel lecture, Robert Millikan noted "Bohr's epoch- making treatment of spectral lines". Other post trilogy Nobelists with direct connections to Bohr are James Frank and Gustav Hertz for their work on electron impact on atoms (1925).

Recently, Heilbron has pointed out[45] that Bohr's strength was in his ability to be critically constructive in navigating through extant literature. As is well known, criticism did not win Bohr any friendship with Thomson when Bohr first arrived at Cambridge. In the trilogy work, showing the strength of his constructive character, Bohr clearly absorbed Nicholson's work and *mutatis mutandis* improved on it. His postulated hypothesis about the energy and the orbital frequency in a stable atom was totally an ad hoc introduction, still, with no quantization. Only with further manipulations, it can be shown that Bohr's hypothesis is in effect equivalent to Nicholson's angular momentum quantization condition.

Although his earlier notes in the memorandum does not show his grasp of the importance of Nicolson's results but by the time of the trilogy papers Bohr not only understands the significance of astronomical spectroscopic lines first introduced by Nicholson, in any quantum description; consequently, he integrated this astronomical perspective with laboratory results to create his own atom. More specifically, Bohr was the first to apply Planck's radiation quantization condition to atomic transitions between energy states, expressing the results in terms of frequencies, not in terms of wave lengths (or inverse wave lengths) as often used by his predecessors such as Balmer, Rydberg and Nicholson etc. Bohr's work on multi-electrons (especially on multi-nuclear systems) did not stand the test of times, but even today his model is an excellent introduction to the subject. Remarkably, it was Nicholson himself who disavowed his own original ownership of the angular momentum quantization idea, by becoming the most vocal opponent, at least, in Britain, of (Bohr's) atomic theory.

As John Ziman has observed[46] scientific pursuit, the seeking of truth, is also as much "the game as competition" [ JZ p 44]and has mentioned its recognition process as "…notable scientific achievement by awards of medals and prizes" [JZ,p45]. By this measure, clearly Bohr's work has gained its place in the community of modern physics amongst academics. As Ziman also noted, there has been a symbiotic interdependence between institutionalized awards and "good science" (Zeeman, p.44).

The most prestigious physics honor today in 2013 is the Nobel Prize. At the same time, physics also has an eminent place in the prize itself, since in his will Alfred Noble identified physics as the first; "[Thus,] [t]he said interest shall be divided into five equal parts, which shall be apportioned as follows: … one part to the person who shall have made the most important discovery or invention within the field of physics ...".  It is arguable as to what comes first, the esteem of the great scientists, or their "inventions" or the prominence of this prize.  Nevertheless, it will be fair to assume that at the time of the first awarding in 1901 Roentgen himself had more name recognition than the prize itself.  Now, in 2013, after 106 prizes have been awarded to a total of 194 Laureates, it is in fact the prize that has accrued in stature. The Nobel organization has morphed into one of the most influential science 'institutions' that Ziman describes[46]. Arguably Nobel's global reach is palpable thru continuing impacts in funding decisions, choices in research directions and academic recruitments.

According to the Nobel web site [http://www.nobelprize.org/nobel_prizes/physics/] the eight most popular Laureates in increasing influential order are – Planck, Marconi, Thomson, Roentgen, Feynman, Bohr, (Marie ) Curie, and Einstein. An asterisk mark in Table (III) indicates direct relevance to the present article either in connection with atoms, vacuum discharge, X-ray, or quantization. From the list of attendees at the first Solvey council (Table I) in 1912-13, it would have been reasonable to expect many of the then (young) leaders were "shoe ins" for a Nobel- it was just a matter of time; for example, Einstein's divorce agreement with Mileva Einstein nee Maric already showed that his prize was almost 'a given'.

**Legacy of Bohr's theory:**

The legacy of Bohr is immense and pervasive, extending beyond the confines of science. It was truly revolutionary for him to be so bold in 1913 to demand that the same quantum principles should apply to physics exactly the same way as to chemistry, hence by implication, to all of nature. Culturally, his vision of electrons as tiny planets orbiting around a nucleus is the signature icon of the nuclear age, as seen in most logos of atomic organizations such as in that of IAEA.

Bohr's model addresses a particular set of 'physical' behaviors of the atom. With this model Bohr obtains clear answers to queries e.g., by what amount is the energy of the atom increased (or decreased) or what is the related frequency of radiation. However, the model does not pretend to describe all physical processes; Most of all, certainly not by what process is the radiation absorbed or emitted. It was true, at that time, many including Einstein and Bohr himself were preoccupied with "process" of how the electron (atom) radiates (or absorbs) quanta of energy and also with the relationship between the frequency of the field and the orbiting frequency of the radiating electron (in the modern respective, it is rather the whole atom that radiates). Throughout his career, Bohr went on to profess the importance of the correspondence principle in relating the quantum results with the classical ones.

Even after the advent of full quantum mechanics, Bohr's ideas continued to be influential for decades. Nicholson started as the early front runner, but in the post-world war 'winner take all' attitude, he has been gamed out. Not only has the attention to Nicholson's works decreased with time, but also even other works in quantization such as 'Wilson-Sommerfeld' condition (rule) today is more familiar as the 'Bohr-Sommerfeld etc.', instead. For instance, a Google Scholar search in August 2013 for "Wilson-Sommerfeld quantization" got 2000 results whereas a similar search performed at the same time for "Bohr-Sommerfeld quantization" generated 8,430 results. Perhaps, this shows another classic case of Mathew effect, "where honors and resources accumulate around who already have them" [p 46, Ziman (2000)]. Buoyed by the 2013 centennial, the iconic role of Bohr is on the rise. Another recent Google Scholar citation search showed that the number of citations for the trilogy Part 1 has grown ten folds to about one hundred per year over the last couple of years. Today, the century old Bohr's model has lost much of its scientific importance, but still remains relevant as a primer for quantum physics.

**Finally Attempting to Answer Heilbron & Kuhn 1969:**

Let us now present a line of reasoning that may answer Heilbron and Kuhn's question posed in their celebrated 1969 article[47] - Genesis of the Bohr Atom. These authors asked "what suddenly turned his [Bohr's] attention … to atom models during June 1912" . Some may insist that there really was nothing sudden. But, H&K were absolutely right; during the short period in question Bohr had made an unexpected change in his research activity. He has found a new muse, "the atom" and would soon produce a spectacularly successful theory about it. Long gone is the minutia of electron theory. Bohr writes (June 19, 1912) to his brother Harald; "Don't talk about it to anybody, for otherwise I couldn't write to you about it so soon… I have taken off a couple of days from the laboratory (this is also a secret)." Oh, Bohr got secrets! What was there to conceal? Wasn't he working on an improved theory of alpha and beta particle scattering, devoting all of his energy and extra time? Isn't doing big calculations on "velocity of electrified particles" justifiable and authorized? Furthermore, he wrote to his brother "also a secret". Implying that there was at least one more 'the original' secret, then why did Bohr not say what was this secret? We reason that the "elephant in the room" was nothing but the inspirational news about the great discovery with X-rays.

The biggest physics news in May-June of 1912 came from Munich with Laue's discovery. News of Laue's discovery arrived to Henry Bragg by second hand via letter from Lars Vagard in late June of 1912. This result transformed X-rays into a potent research tool. Furthermore, the Laue pictures were the first visual proof for the physical existence of atoms. As Lawrence Bragg stated, X-rays revealed, for the first time with thousand times better resolution than visible light, "the actual atoms in their permanent state" [N Bohr, P164 Phil Mag S.6.vol 26.no151,July1913] are in perfect geometric arrangement of "absolutely fixed dimensions"[p 164]. The Braggs were electrified into immediate action, and quickly published a series of remarkable papers one after another. They created various new fields in science and within two years became Nobel Prize winners!

Is it not likely that folks in Manchester would have also received the news of Laue spots directly from Sommerfeld in Munich and possibly even earlier than Henry Bragg?

At this time, there was another talented young researcher in Manchester. H.G.J. Moseley's research topic is described by Heilborn [John L. Heilbron, The Work of H. G. J. Moseley, History of Science Society, vol 57, No 3 pp-336-364 (1966) to be "His [Moseley's] choice fell on a field which had not been cultivated at Manchester, but which at that moment was perhaps the most exciting subject in physics – X-rays, … in July 1912, by the discovery [of] … diffraction pattern….Moseley and Darwin were [also] interested in X-rays as a fundamental problem. … [F]inally the master [Rutherford] was persuaded to let them try, and sometime in the fall of 1912 Moseley set off for Leeds to learn the mysterious art of X-ray experimentation". It is clear that Moseley was excited enough by the Munich discovery to start doing his own X-ray research!

Cliché's abound- seeing is believing or a picture is worth a thousand words etc., really after Laue's pictures, there can be no more doubts about the physical reality and hence stability of atoms. It would be natural for the folks in Rutherford's laboratory to inform Bohr about this discovery and exclaim a declaration for instance to the effect – *Eureka! I have seen Laue's photographs, they show tiny atoms, with absolute certainty fixed, in perfect order. Natural atoms are resistant to change, no problems with stability etc, etc.* A remark of this nature can be the jolt that inspired Bohr to get busy with the atom theory in the summer of 1912.

Unfortunately, we are aware of no known published record of such remarks to Bohr by Rutherford, by Moseley or by anyone else. Also, Bohr himself never mentioned receiving such intimation, either. However there is one strong indication. In Part-1 of trilogy as he sets the stage to discuss the stability of atoms, Bohr invokes "an atomic system occurring in nature […], the actual atoms in their permanent state […] have absolutely fixed dimensions" [N Bohr, P.4 Phil Mag S.6.vol 26.no151 1 July1913] . We ask in 1912-1913 how one would know of "absolutely fixed" (atomic) anything? The best one could get would have been from Laue pictures. Furthermore, in Part-2, Bohr notes "by the interference observed in recent experiment on diffraction of Rontgen rays in crystals" [p 500, N Bohr, Phil Mag S.6.vol 26.,476]   So, here in the middle of the trilogy in print, Bohr acknowledges that he definitely knew about X-ray and crystal lattice. Also, his mentioning of both interference (a term used by the Braggs) and diffraction (by Laue) of X-ray shows his full familiarity with the works of both Laue and the Braggs. Remarkably, however, no single bibliographic reference is given and Bohr is totally silent about Laue's name. It will be interesting to locate original correspondence on the subject between Rutherford and the others at or around Manchester from that period. Exactly when Bohr learned about Laue's discovery and how much he found it inspiring can be questions for future researchers. Perhaps in the archives in Manchester and in Cambridge or perhaps in Rutherford's collections, one might be able to find another "way of being" to this story. In summary, at least circumstantially Bohr was fortunate enough to be at the right place at the right time, he also had the means, the desire and support he needed to complete his life's biggest achievement. We tried to show "Laue-news" could have been the "secret" trigger that got him started in the quantum atom race, overshadowing the early starter Nicholson. Bohr left an inspiring legacy for generations of Nobel Laureates.

**Where we wearily draw to a close:**

In this article, we have outlined a story line tracing the development on the quantum theory for atoms with a number of distinct subplots; X-rays provided the first direct evidence about the structure of matter and into the atom itself. Atomic structure of matter unveiled itself unexpectedly in Sommerfeld's laboratory by the instigation of his optics lecturer Max v Laue. Despite initial misunderstandings, Laue's genius came thru in his prompt recognition of Bragg's (correct) formulation; Laue and the Braggs were recognized with physics Nobel prizes two years in a row, 1914 & 1915. In addition to prompting the Braggs and Moseley, the news of Laue's discovery in June 1912 activated Bohr's impetus dash into quantum atoms. The Laue-Bragg nexus is well documented and confirmed, unfortunately that to Bohr is absent.  However it is clear from his publications that at least by the time of the trilogy Bohr was aware of Laue's discovery. In this subplot, there was also an early leader John Nicholson who had provided the correct reason for atomic stability and introduced the quantization of angular momentum in print a year ahead of Bohr. Right after the appearance of the trilogy, Nicholson disagreed with Bohr's theory. As of 2013 Bohr continues to be the iconic figure in the history of the quantum.

Historian John H. Arnold has noted[48] "…'Origins' are simply where we pick up the story … 'Outcomes' are where we wearily draw to a close" [P 91,Arnold (2000)] and also "history allows us to demur … there have always been many courses of action" [P 122, Arnold (2000)]. John Arnold is correct and often times there are "many ways of being"(p.122). Here is a story for the first quantum atom, one 'way of being', and a plausible answer to Heilborn and Kuhn's question from 1969 is hereby presented. We welcome other ways of being of the past that waits in the future.

**Acknowledgements:**

One of us TD would like to thanks his former USC- Nanocenter colleague John M Thomas for sharing his preprints regarding the Laue –Bragg centenary and also his collaborator Ronald Edge (emeritus) for personal reflections from the days at Cavendish under WLB. TD also thanks Department of Astronomy and Space Sciences at Sejong University, Seoul, Korea as well as both Department of Physics and Department of Materials Sciences at University of Incheon, Incheon, Korea for partial support during his visits last year when the work for this manuscript was started. Especially, TD would like to express his appreciation for Jae Yoon Park for the warm hospitality at the University of Incheon, Incheon, Korea.

**References:**

**A:** [http://www-outreach.phy.cam.ac.uk/camphy/positiverays/positiverays3_1.htm ] Other Nobel Prize worthy photos includes Laue spots, cloud chamber tracks including the discovery of nuclear disintegration of nitrogen, picture of the first atomic beam splitting by Stern-Gerlach and bubble chamber tracks. Visual recording remained central in a number of basic sciences well into the middle of the twentieth century, for instance, such as in auto radiography and photo emulsion techniques. It is well known that until the advent of electronic photo detectors, much of observational astronomy has been dependent on the photography.

**Table-I: List of invitees to the 1<sup>st</sup> Solvay conference (1911)**

        H. A. Lorentz (Leiden), as Chairman.

**From Germany**
W. Nernst (Berlin)
M. Planck (Berlin)*
H. Rubens (Berlin)
A. Sommerfeld (München)
W. Wien (Würtzburg)**
E. Warburg (Charlottenburg).

**From England**
Lord Rayleigh (London, did not attend)**
J. H. Jeans (Cambridge)
E. Rutherford (Manchester)**

**From France**
M. Brillouin (Paris)
Madame Curie (Paris)**
P. Langevin (Paris)
J. Perrin (Paris)*
H. Poincaré (Paris)

**From Austria**
A. Einstein (Prag)*
F. Hasenöhrl (Vienna)

**From Holland**
H. Kamerlingh Onnes (Leiden)*
J. D. van der Waals (Amsterdam)**

**From Denmark**
M. Knudsen (Copenhagen)

Two (**) indicate Nobel Laureates with award prior to the first Solvey council, others with one asterisk are those received the prize after 1911.

**Table –II: List of authors with number of citations referenced in the trilogy**

| Author Name: | Part-I | Part-II | Part-III |
|---|---|---|---|
| C.G.Barkla* | x | 1 | x |
| N.Bjerrum | x | x | 2 |
| N. Bohr* | 1 | 1 | 1 |
| A v d Broek | x | 1 | x |
| A H Bucherer | x | 1 | x |
| C & M Cuthbertson | x | 1 | 1 |
| Verh Deutsch | 1 | 1 | x |
| A. Einstein* | 4 | x | x |
| K Fajaus | x | 2 | x |
| A Fowler | 1 | x | x |
| J Franck , u G Hertz et al* | x | 2 | x |
| Geiger & Marsden | 1 | 1 | x |
| P. Gemlin | x | 1 | x |
| F. Haber & Verh deutsch | 1 | x | x |
| A Hass | 1 | x | x |
| G v Hevesy* | x | 1 | x |
| H.Kayser | x | x | 1 |
| I.Langmuir* | x | x | 1 |
| F A Lindermann & Verh Deutsch | 1 | x | x |
| R.A.Milikan* | x | 1 | x |
| J W Nicholson | 7 | 2 | x |
| F Paschen | 1 | x | x |
| E C Pickering | 2 | x | x |
| M Planck* | 3 | x | x |
| W Ritz | 1 | x | x |
| A.S.Russel & R. Rossi | x | 3 | x |
| E Rutherford** | 3 | 2 | x |
| A Schidlof | 1 | x | x |
| F Soddy | x | 1 | x |
| JJ Thomson** | 4 | 1 | 2 |
| E. Warbur, G. Laithauser et al | x | 1 | x |
| E Wertheimer | 1 | x | x |
| R. Whiddington | x | 1 | x |
| R Wood | 1 | x | x |
| **Total** | **35** | **24** | **8** |

Table –II, The trilogy referenced sixty seven works by a thirty three authors. The names marked with two (**) indicate Nobel Laureates with award prior to the trilogy, the seven others with one asterisk are researchers who's contributions were recognized after 1913.

**Table-III: List of Physics Nobel Prize awards since its inception in 1901 to 1930**

1930 -Sir Chandrasekhara Venkata Raman
1929-Prince Louis-Victor Pierre Raymond de Broglie
1928-Owen Willans Richardson
1927*-Arthur Holly Compton, Charles Thomson Rees Wilson
1926*-Jean Baptiste Perrin
1925*-James Franck, Gustav Ludwig Hertz
1924*-Karl Manne Georg Siegbahn
1923*-Robert Andrews Millikan
1922*-Niels Henrik David Bohr
1921*-Albert Einstein
1920-Charles Edouard Guillaume
1919*-Johannes Stark
1918*-Max Karl Ernst Ludwig Planck
1917*-Charles Glover Barkla
1916-No Nobel Prize was awarded this year. The prize money was allocated to the Special Fund of this prize section.
1915*-Sir William Henry Bragg, William Lawrence Bragg
1914*-Max von Laue
1913-Heike Kamerlingh Onnes
1912-Nils Gustaf Dalén
1911*-Wilhelm Wien
1910-Johannes Diderik van der Waals
1909-Guglielmo Marconi, Karl Ferdinand Braun
1908-Gabriel Lippmann
1907-Albert Abraham Michelson
1906*-Joseph John Thomson
1905*-Philipp Eduard Anton von Lenard
1904-Lord Rayleigh (John William Strutt)
1903*-Antoine Henri Becquerel, Pierre Curie, Marie Curie, née Sklodowska
1902-Hendrik Antoon Lorentz, Pieter Zeeman
1901*-Wilhelm Conrad Röntgen

The asterisk (*) marks indicate contributions in atoms, vacuum discharge, X-ray, or quantization and hence are relevant to the present article.

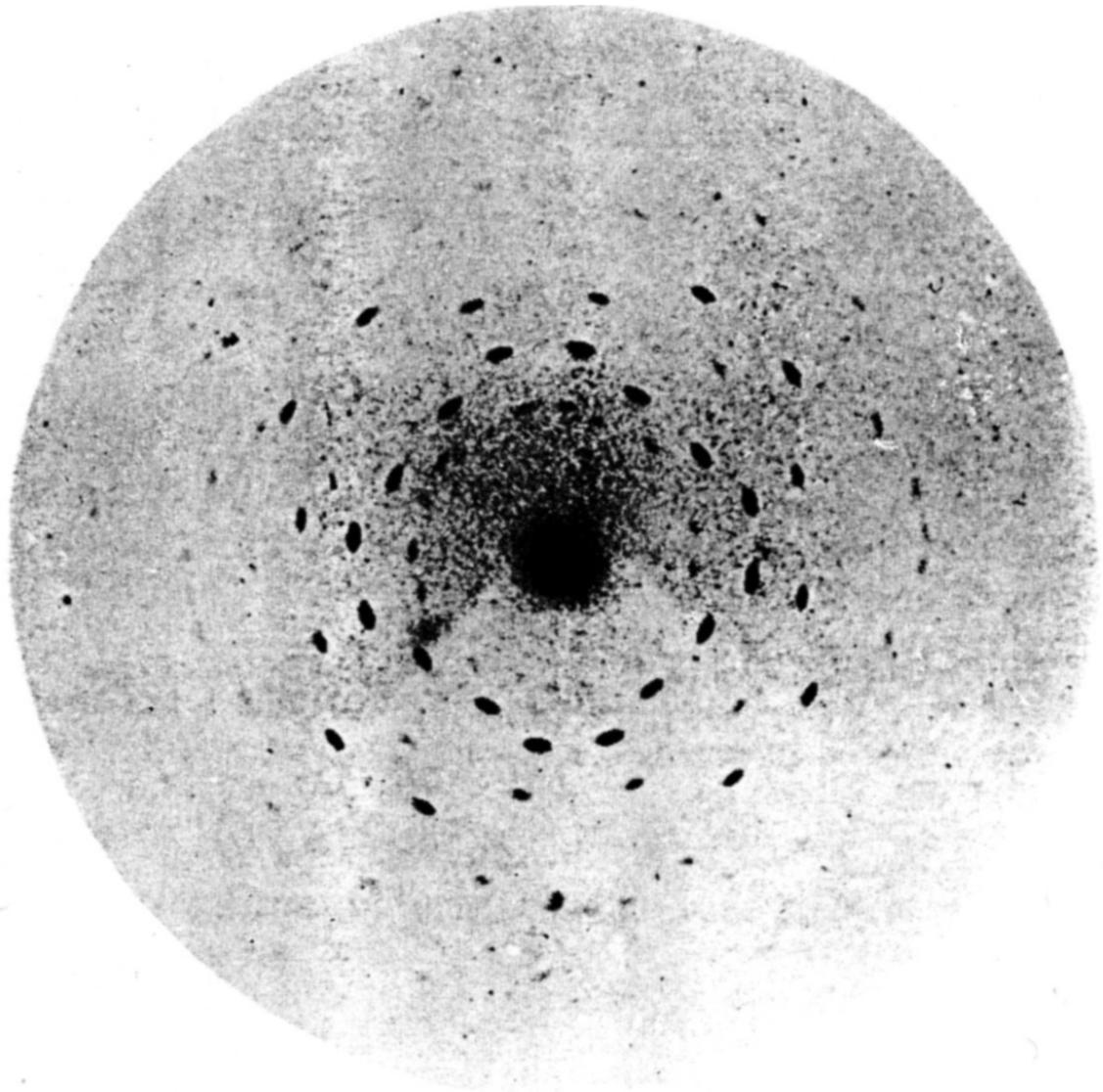

**Figure 1:** An early image of a photo-plate showing 'Laue spots' taken in Munich in 1912. As described in the text Prof. Henry Braggs at Leeds and son Lawrence Bragg at Cambridge as well as Rutherford's graduate student Moseley at Manchester were inspired by Laue's discovery to experimentally investigate atomic properties. We reason Rutherford's post-doctoral assistant Bohr also at Manchester in the summer of 1912 was similarly inspired to develop the quantum theory of the atom.

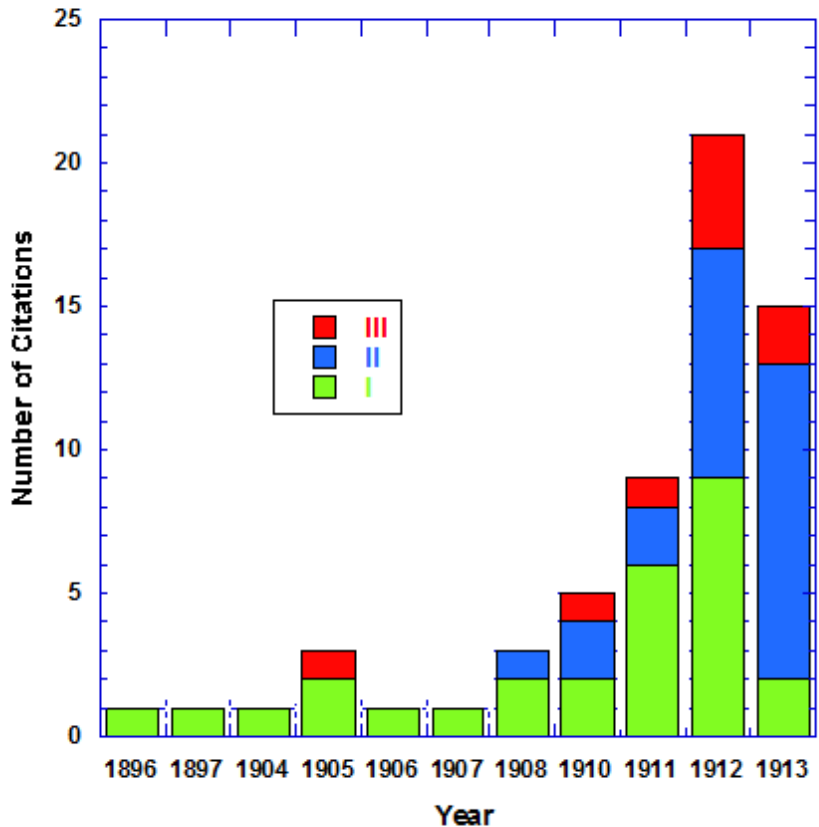

**Figure 2:** A bar graph of Bohr's citations in the three parts of his now famous trilogy papers published in the Philosophical magazine in 1913. The axes represent the total number of counts of references from each year. As described in the text the publication dates of the cited works range from 1896 to 1913. The three papers cited a total of sixty five distinct (numbered) references from thirty one authors. The first paper (part-I) had the most citations, thirty five, the second paper had twenty four and the last one only seven citations.